# QUICKER REACTION, LOWER VARIABILITY: THE EFFECT OF TRANSIENT TIME IN FLOW VARIABILITY OF PROJECT-DRIVEN PRODUCTION

Ricardo Antunes[1], Vicente González[2], and Kenneth Walsh[3]

## ABSTRACT

Based on the knowledge of dynamic systems, the shorter the transient response, or the faster a system reaches the steady-state after the introduction of the change, the smaller will be the output variability. In lean manufacturing, the principle of reducing set-up times has the same purpose: reduce the transient time and improve production flow. Analogously, the analysis of the transient response of project-driven systems may provide crucial information about how fast these systems react to a change and how that change affects their production output. Although some studies have investigated flow variability in projects, few have looked at variability from the perspective that the transient state represents the changeovers on project-driven production systems and how the transient state affects the process' flow variability. The purpose of this study is to investigate the effect of changes in project-driven production systems from a conceptual point of view, furthermore, measuring and correlating the transient response of five cases to their flow variability. Results showed a proportional relationship between the percentile transient time and flow variability of a process. That means that the quicker the production system reacts to change; the less the distress in the production output, consequently, lower levels of flow variability. As practical implications, lean practices focusing on reducing set-up times (transient time) can have their effects measured on project-driven production flow.

---

[1] Ph.D. candidate, Department of Civil and Environmental Engineering - University of Auckland, New Zealand, rsan640@aucklanduni.ac.nz
[2] Senior Lecturer, Department of Civil and Environmental Engineering - University of Auckland, New Zealand, v.gonzalez@auckland.ac.nz
[3] Dean, SDSU-Georgia, San Diego State University, Tbilisi, Georgia, kwalsh@mail.sdsu.edu

73

# INTRODUCTION

The importance of the time spent in production changeovers has been well-known for a long time (Taylor, 1911), as much as structured approaches to reduce such time (Gilbreth, 1911). However, it was later with the development of 'Just-In-Time' that progress was made to lessen the changeover time. Observations of how to reduce the time spent on the exchange of dies on automotive pressing machines resulted in a structured methodology capable of bringing down the time spent in changeovers from hours to minutes (Shingō, 1985). Widely applied in the manufacturing till this day, the Single Minute Exchange of Dies (SMED) consists of seven basic steps. The steps are:

(I) observe and measure the current methodology;

(II) separate external of internal activities;

(III) transform internal activities into external ones;

(IV) simplify remaining internal activities;

(V) make the external activities more efficient;

(VI) standardize the new procedure; and

(VII) repeat the method for further improvement

## THE 3 UPS OF CHANGEOVER

A process changeover (Figure 1) consists of three 'ups':

### Cleanup

Cleanup is the removal of previous product, materials, components, and/or residuals from the production line or site. It may range from minor tasks such as cleaning after a painting job has finished to major work such as a tower crane disassembling. '5S' practices perform a significant role in the cleanup stage, simply because if there is less to clean, it can be done faster (Womack, Jones, and Roos, 1991). Hence, keep a clean and organized site supports a quicker cleaning. Ideally, the cleanup finishes with the production output. It means that once a production output reaches zero, the site is clean. Accordingly, the cleanup stage can be measurable. It starts when the input of the production system ceases. In the best case scenario, the cleanup finishes when the process output reaches zero, *i.e.*, there is nothing else to be produced or cleaned. Otherwise, the cleanup finishes when everything needed in the process is removed.

### Setup

Setup is the group of activities of converting the site to run a new process. The conversion requires adjusting or parametrizing equipment to match the requirements of the next production process or by replacing non-adjustable equipment. Usually, it involves a combination of both. For that reason, the resources are applied for preparing the site for the process while the production process stands still waiting to start. This situation is the *muda* of waiting (Ohno, 1988), or simply waste. The setup stage is utterly unproductive; it adds no value. Therefore, lean practices aim to 'zero setups' or 'eliminate setup'. The setup is also measurable. It begins when cleanup finishes and ends at the production kick-off.

**Startup**

Startup (transient) is the time immediately after a process kick-off until the full process operation (steady-state). The initial moments involve 'learning' and fine-tuning the equipment and the pace of work. In this stage, jams and stoppages are frequent causing defective products and variations in the production output (Shingō, 1985). The production system often underperforms at the setup stage in comparison to the production at steady-state.

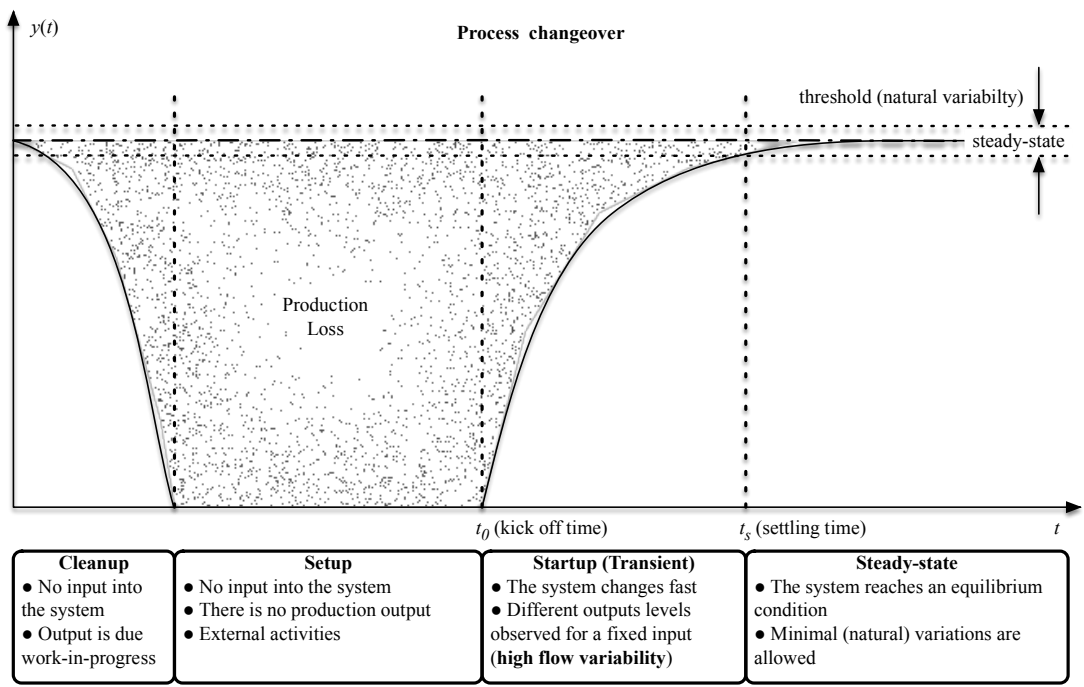

*Figure 1 - Process changeover*

## SMED AND CONSTRUCTION

The main focus of SMED is the transformation of internal activities of the setup stage to external activities. In manufacturing, an internal activity is any operation that can only be performed if the machine is shut down (for instance, attaching or removing the dies). An external activity can be executed when the machine is running (Cakmakci, 2008). In project management terms, internal activities are in the critical path, while external activities are parallel to the critical path. Hence, in a project-driven process, the application of SMED means removing activities that are not hardwired to the critical path and executing them in parallel, furthermore, resulting in a compressed critical path. In the end, SMED practices in project management can be seen as a method for fast tracking the project schedule.

The concept and benefits of SMED are well known in the construction industry, especially within the lean construction community. An example of the use of SMED practices in construction is the offsite fabrication (Gibb, 1999). On offsite fabrication building tasks are performed externally to the site and what is left is a reduced number of assembling activities to be performed onsite. Automation is another example of the use of SMED, in particular, the application of techniques to eliminate adjustments and improve mechanization. Construction processes are flexible, constituted by a workforce, machinery and equipment with relatively general purpose (Hayes and Wheelwright, 1979, p. 134). Such flexibility favors the adaptation of existing resources to new processes over resources replacement in the setup stage. However, there is still the 'adjusting or parametrizing equipment' to do. Automatic parametric machinery may reduce the time of setup times by performing the conversion faster and improve startup because it may reduce the possibility of human error while adjusting the machinery. Although the use of offsite fabrication and automation in construction are strongly related to SMED and widely present in the construction industry, little is the discussion about the structured application of SMED as performed in manufacturing. Even though, the implementation of SMED in construction is likely to be easier than in manufacturing based on general purpose equipment and workforce in the construction industry.

As all continuous improvement techniques, SMED requires measurement, control, comparison and benchmarking. As measurements, SMED usually utilizes the estimator of Process Performance Index (assuming that the process output is approximately normally distributed), $C_{pk}$, and Process Performance, $P_p$, values for judging a process, whether it is capable of improvement or not (Cakmakci, 2008). $C_{pk}$ is the result of the upper specification limit minus the mean of the output divided by three times the output sample standard deviation; or, the mean of the output minus the lower specification limit divided by three times the output sample standard deviation. Whatever shows the minimum value (Montgomery, 2009, p. 355). The $P_p$ is given by the difference of upper specification limit and lower specification limit (here plus and minus two percent around the mean, matching the threshold limits) divided by six times the standard deviation of output sample (Montgomery, 2009, p. 363).

Those formulas provide unique values of $C_{pk}$ and $P_p$ that can be used in the judgment, what works fine for manufacturing. Construction is a different story. The difficulty in judging a construction process is due to the short run (batch size) of its production (Antunes and Gonzalez, 2015). In other words, there is not a long enough—sometimes any—steady-state to produce useful data (normal distribution around the steady-state value) to use $C_{pk}$ and $P_p$. The highly variable and long transient state of project-driven processes disrupt the accuracy of the values given by $C_{pk}$ and $P_p$. Because they end up accounting for variations in transient—and consequently setup—stages rather than the variations at steady-state. Another difficulty is in defining what are the upper and lower limits of variation once project management problems may have several suitable solutions.

## FLOW VARIABILITY

In a process chain (Figure 2), the output of a process is the input of another, consequently, establishing a flow. When variations of the output of the process $i$ affects the input, and/or behavior of the following process, $i+1$, this is called flow variability.

How much the output variation of the process *i* affects the process *i*+1, depends on two factors. One is the coefficient of variation of the arrival rate of the process *i*+i, $c_a$. In a process chain, without yield less or rework, the arrival rate of the process *i*+1 is equal to of the departure rate of the process *i*, as well as the coefficient of variations, $c_{d(i)} = c_{a(i+1)}$. That is known as the conservation of material (Hopp and Spearman, 2001, p. 253).

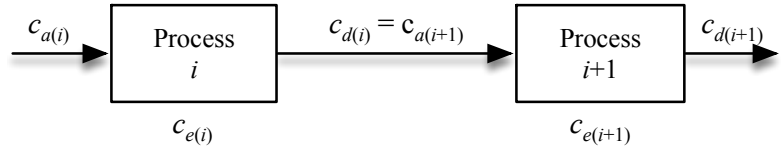

*Figure 2: Propagation of variability between processes. Adapted from Factory Physics: Foundations of Manufacturing Management, 2nd ed. (p. 262), by Hopp, W. J., and Spearman, M. L., 2001, Irwin McGraw-Hill.*

The second factor is the utilization, *u*, of the process *i*+1. *u* values close to one indicate a process almost always busy, on the other hand, values close to zero points out a process nearly always idle. Since *u* is likely to assume values between zero and one the output variation of the process, $c_{d(i+1)}$, is given by Equation 1. Accordingly, if the output of the upstream process *i* is highly variable, the output of the downstream process *i*+1 is also highly variable (Hopp and Spearman, 2001, p. 261).

$$c^2_{d(i+1)} = u^2 c^2_{e(i+1)} + (1 - u^2) c^2_{a(1+i)}$$

*Equation 1: Coefficient of the departure of the downstream process*

### MEASURING PROCESS' TRANSIENT

The direct method of measuring a process' transient is to compute the amount of output at regular intervals of time, *e.g.*, 'Time and Motion' (Taylor, 1911). Nevertheless, to calculate the transient time, $t_s$, the process must reach the steady-state. Hence, the data collection must proceed until the steady-state is reached and is undisputed that the process is in this state. Such procedure seems impractical as a non-automated task. Because, that means to utilize a workforce to monitor, measure and count the production output regularly in short periods of time. Even, after all, it may be impossible obtaining the process' transient time because construction processes often do not reach the steady-state.

### PRODUCTIVITY FUNCTION METHOD

A productivity function, *P(t)*, represents the relation between the output function, *Y(t)*, and the input function, *U(t)*, of a project-driven production process, *Y(t) = P(t)*U(t)*, * symbolizes the convolution operator. Approaching the production process as a dynamic system the productivity function accounts for the transient and steady/unsteady-state (Antunes, González, and Walsh, 2015). The transient time is given by the transient analysis (Figure 2) of a processes' productivity function. The transient time, $t_s$, is the time the output takes to reach the steady-state value, or a threshold around the steady-state value (usually, –+2%) from the moment a unitary step input is applied, $t_0$. The step input acts as an off-on switch, *e.g.*, a light switch, which the input changes instantaneously from zero to one. The change in the input provokes the reaction of the system that tries to adapt as fast as possible (the bulb light filament warms once there is an electric current). Later on, the output tends to a constant value when $t \rightarrow \infty$ (the filament reached a temperature in which it produces a steady amount of light). The percentile reaction time is obtained by dividing the process transient time, $t_s$, by the total process time, $t_t$.

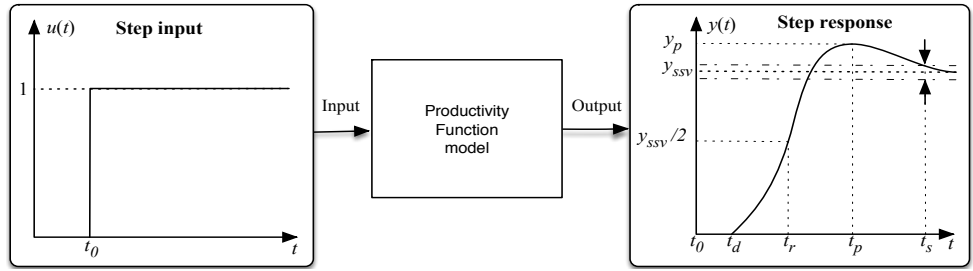

*Figure 2 - Transient analysis*

Some studies have investigated flow variability in projects. However, few have investigated the relation and effects of the transient state on flow variability of project-driven production systems. This study aims to examine, from a conceptual standpoint, the effect that changes in project-driven production systems have on their flow variability. Furthermore, measure and correlate the transient response of five cases to their flow variability.

## METHOD

This research analyzed five cases with different sample sizes, and the processes compile various activities in construction. The cases are:

Case 1: drilling an offshore oil well (Antunes et al., 2015),

Case 2: wall assembling (Abdel-Razek, Elshakour, and Abdel-Hamid, 2007),

Case 3: setting formwork for slabs,

Case 4: group of activities (foundation excavation and backfill) from a housing project, and

Case 5: wall plastering (González, Alarcón, Maturana, Mundaca, and Bustamante, 2010).

The commonality among the cases is that they configure a system, *i.e.*, they are constituted by input, transformation process, and output.

First, the process output variation is measured. Process Performance Index, $C_{pk}$, and Process Performance, $P_p$, are obtained for the cases and shown in Table 1. A third measurement of output variation is given by the coefficient of variation (Hopp and Spearman, 2001, p. 252) of the departure times the process $i$, $c_{d(i)}$. That is the result of the ratio of standard deviation of the time between departures (standard deviation of the output) and the mean departure rate (output mean). The results are also shown in Table 1.

Before a productivity function can be obtained, an accuracy benchmark must be set. Thus, a first-degree polynomial model (FDP), $y(t) = \alpha u(t)$, is estimated using the regression analysis and the goodness of fit sets the benchmark. The productivity functions are then estimated by trial and error approach (Antunes et al., 2015). Later, the transient time, $t_s$, is obtained in the transient analysis (Figure 2) and the percentile reaction time can be determined.

# RESULTS

Table 1 displays the transient analysis results given by the production function and the flow variability obtained by statistical means. Rather than sorted by case number, Table 1 was sorted by ascending 'percentile reaction time' to ease visualization and correlation between the two methods. Table 1 shows that as the percentile reaction time values increase the values of $C_{pk}$ and $P_p$ decrease, indicating a lower estimated probability of a process' output being within the limits. Additionally, Table 1 shows as the percentile reaction time values increase the coefficient of variation of the processes also increase meaning that the output variation increases. Consequently, the flow variability (Equation 1) also increases.

It is important to mention that the result of Case 2 has the highest percentile reaction time of the cases. The percentile reaction time is over 100% indicating that even if the processes input was kept constant during the process life spam the output would not reach the steady-state, or that process production run is too short for reaching the steady-state. Hence, it can be said that the process described in Case 2 is unsteady. The second point on Case 2 is that it is the only one with 'moderate variability' ($0.75 < c_d < 1.33$). Actual process times are usually 'low variability', $c_d < 0.75$. For a system with workload evenly distributed among the resources, it means that the process operates around the practical worst case (Hopp and Spearman, 2001, p. 232) with characteristics of processes with short adjustments (Hopp and Spearman, 2001, p. 254).

*Table 1 - Flow variability and percentile reaction time*

| Case | Transient (settling) time $t_s$ | Total process time $t_t$ | Percentile reaction time $t_s / t_t$ | *Process capability index* $C_{pk}$ | *Process performance* $P_p$ | Std. dev. departures $\sigma_d$ | Departure rate $r_d$ | Coefficient of variation $c_d = \sigma_d / r_d$ |
|---|---|---|---|---|---|---|---|---|
| 1 | 4.67 | 184 | 2.54% | 0.2438 | 0.5354 | 95.22 | 3481.40 | 0.0273 |
| 4 | 13.11 | 210 | 6.24% | 0.0144 | 0.0344 | 14.89 | 32.12 | 0.4634 |
| 5 | 1.23 | 18 | 6.86% | 0.0110 | 0.0260 | 278.44 | 458.68 | 0.6070 |
| 3 | 1.82 | 8 | 22.71% | 0.0108 | 0.0258 | 315.13 | 510.27 | 0.6176 |
| 2 | 55.85 | 20 | **279.26%** | 0.0085 | 0.0204 | 27.81 | 35.55 | **0.7824** |

$t_s$ – transient (settling) time
$t_p$ – total process time
$r_d$ – departure rate (output mean the output, *i.e.*, $\bar{y}$)
$\sigma_d$ – standard deviation of the time between departures (standard deviation of the output, i.e., $\sigma_y$)
$c_d$ – coefficient of variation of the departure times ($c_d, = \sigma_d / r_d$, *i.e.*, CV = $\sigma_y / \bar{y}$)
$C_{pk}$ – process capability index
$P_p$ – process performance

Figure 3 shows the step response of production functions shown in Table 2, Cases 1, 4, 5, 3, and 2.

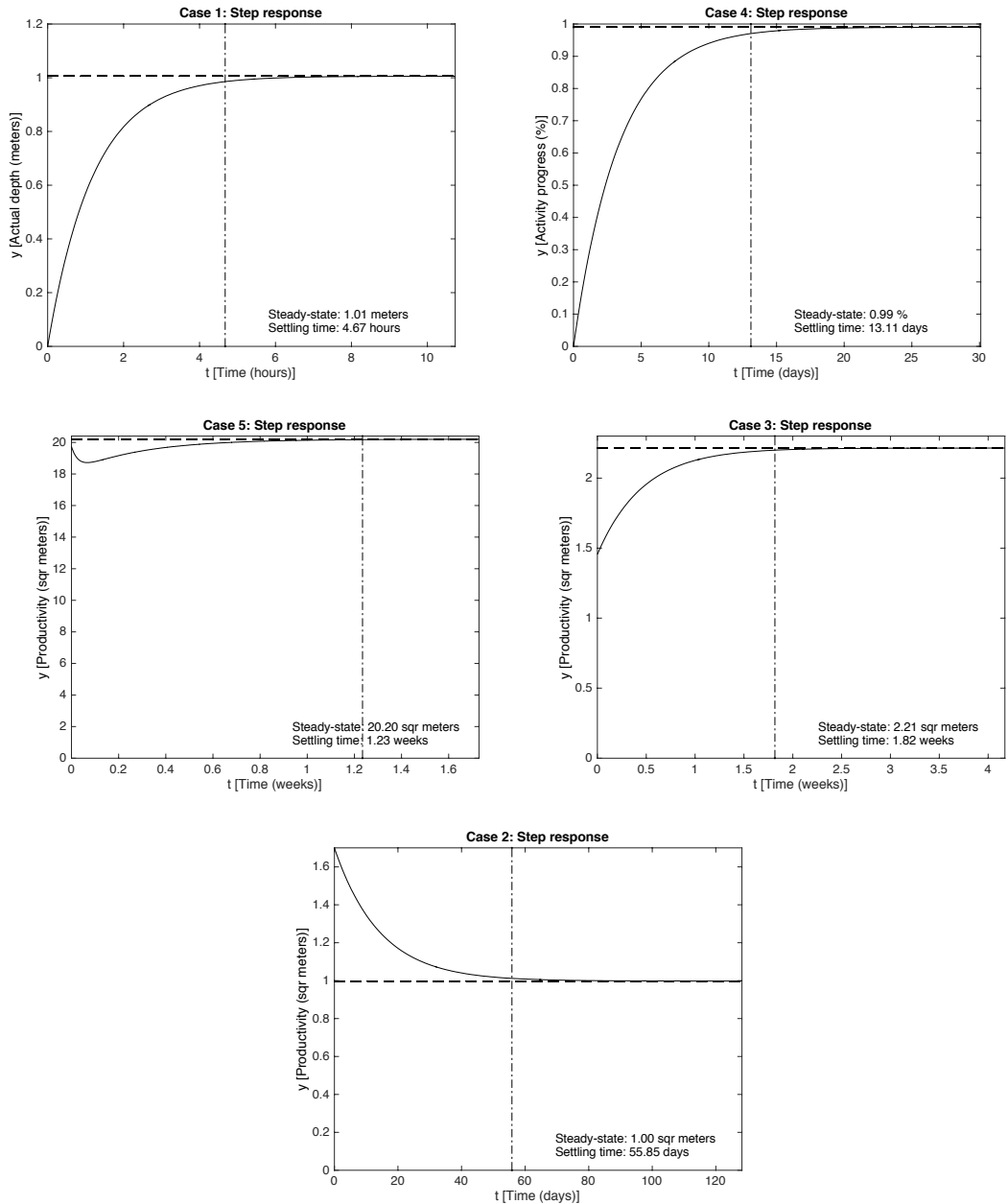

*Figure 3 - Transient analysis of Case 1, Case 4, Case 5, Case 3, and Case 2*

Table 2 shows the production functions, in the time domain, obtained in for the cases and used in the transient analysis, sorted in the same order as Table 1.

*Table 2 - Productivity Functions*

| Case | Productivity Function |
|---|---|
| 1 | $P_1(t) = 0.8417 e^{0.8369t}$ |
| 4 | $P_4(t) = 0.2955 e^{0.2984t}$ |
| 5 | $P_5(t) = 7.142275814 e^{3.444656802t} + 19.74\ (t)\ 54.39107581 e^{33.6753432t}$ |
| 3 | $P_3(t) = 1.455\ (t) + 1.635385 e^{2.153t}$ |
| 2 | $P_2(t) = 1.699\ (t)\ 0.04910796 e^{0.07004t}$ |

# CONCLUSION

Results show a proportional relationship between the percentile transient time and flow variability of a process, confirmed by the coefficient of variation and process capability index calculations. These findings thus lend support that the quicker the production system reacts to change; the less is the distress in the production output, consequently, lower levels of flow variability. The findings align with what is known about dynamic systems and operations management. In manufacturing, the larger the batch, the more efficient the production process becomes. A larger batch requires a longer run time, hence the more irrelevant the setup time becomes when compared to the total run time. The same behavior was observed in construction when calculated the processes' Percentile reaction time using productivity functions. However, in project-driven processes increase the batch size and run times are not desired. Increased batch sizes imply producing more than what is needed: scope creep. That is the *muda* of over-production. Prolonged run times translates into extended activities duration. If the scope is constant the work should be done at a slower peace: decrease of productivity. Hence, the likely option is to reduce the time spent on the startup (transient time). In this fashion, productivity functions may provide a way to measure, visualize and compare the transient state of project-driven processes. Reliance on this method must be tempered, however, because the number of cases analyzed was small. Therefore, it would be beneficial to replicate this method in additional cases of project-driven systems in construction. As practical implications, the understanding of the effects of the transient state on the process variability may induce practitioners to re-evaluate the application of some Lean practices in construction. For instance, SMED practices that focus on reducing set-up times (transient time) can have their effects measured on project-driven production flow supporting a quantitative and structured application of this method.